\documentclass[smallextended]{svjour3}       % onecolumn (second format)
\smartqed  % flush right qed marks, e.g. at end of proof
\usepackage{graphicx}
\usepackage{amssymb}
\usepackage[latin1]{inputenc}
\usepackage{amsmath}
\usepackage{fontenc}
\usepackage{amsfonts}
\usepackage{dcolumn}
\usepackage{bm}

\begin{document}

\title{{\Large Effective spectrum width of the synchrotron radiation}}

\author{V. G. Bagrov \and D. M. Gitman \and A. D. Levin \and A. S. Loginov \and A. D. Saprykin}
\institute{V. G. Bagrov \at Department of Physics, Tomsk State University, Russia; Department of Physics, Tomsk State University, Russia; Institute of High Current Electronics, SB RAS, Tomsk, Russia\\
 \email{bagrov@phys.tsu.ru}
 \and D. M. Gitman \at
Department of Physics, Tomsk State University, Russia;
Institute of Physics, University of São Paulo, Brazil;
P.N.Lebedev Physical Institute, Russia;\\
\email{gitman@if.usp.br}
\and A. D. Levin \at
Institute of Physics, University of São Paulo, Brazil;\\
\email{alevin@if.usp.br}
\and A. S. Loginov \and A. D. Saprykin \at
Department of Physics, Tomsk State University, Russia;
}
\date{\today}

\maketitle

\begin{abstract}
For an exact quantitative description of spectral properties of synchrotron
radiation (SR), the concept of effective width of the spectrum is
introduced. In the most interesting case, which corresponds to the
ultrarelativistic limit of SR, the effective width of the spectrum is
calculated for the polarization components, and new physically important
quantitative information on the structure of spectral distributions is
obtained. For the first time, the spectral distribution for the circular
polarization component of the SR for the upper half-space is obtained within
classical theory. 
\end{abstract}

\section{Introduction}

The theory of synchrotron radiation (SR) is currently a quite well-developed
branch of theoretical physics. Its basic elements have been stated in books
(e.g., \cite{1,2,3,4,5}) and numerous articles. As one of the most
physically important features of SR, one should mention a high polarization
degree of radiation and a unique structure of spectral distribution in the
ultrarelativistic limit. All the theoretically predicted properties of SR
have been confirmed experimentally.

The development of the SR theory makes it possible not only to predict
qualitatively the peculiar features of radiation, but also to propose exact
quantitative characteristics of physically important properties.

For instance, the high polarization degree of SR was predicted qualitatively
by theory more than half a century ago (see, e.g., \cite{1}), and the linear
polarization was given exact quantitative characteristics; however, it was
not until much later that it proved to be possible to obtain exact
quantitative characteristics for circular polarization \cite{6,7}.

In this paper, we propose new exact quantitative characteristics of the SR
spectral distribution, the effective width of the spectrum. We demonstrate
how to calculate this quantity theoretically in the most interesting case,
which corresponds to the ultrarelativistic limit of SR, and what physically
important information can be obtained using this quantity. We demonstrate
that for spectral distributions of SR the well-known characteristics - the
half-width of the spectrum is less informative than the proposed effective
width. We place in the Appendix all necessary formulas for our purposes that
describe the spectral-angular distribution of the SR.
Our consideration is performed in the framework of the classical theory of SR. It is well known that such a theory is valid with high accuracy for accessible parameters of the electron beam in modern accelerators and storage rings. However, for SR in cosmic space  quantum effects can be quite essential (see . Ref. \cite{8}) and thus can significantly change classical results (which may be the subject of a separate study).

\section{The effective width of the spectrum of SR polarization components}

Let us define the effective width of the spectrum $\Lambda_{s}(\beta)$ as
the minimum spectral range that accounts for at least half of the total
radiated power of a given polarization component,%
\begin{equation}
\Lambda_{s}(\beta)=\nu_{s}^{(2)}(\beta)-\nu_{s}^{(1)}(\beta)+1\,. 
\label{13}
\end{equation}
The harmonics $\nu_{s}^{(1)}(\beta)$ and $\nu_{s}^{(2)}(\beta)$ determine
the beginning and the end of this minimum spectral range and are determined
as follows:

Let us introduce the quantities%
\begin{equation}
\widetilde{\Phi}_{s}(\beta;\,\nu^{(1)},\,\nu^{(2)})=\sum_{\nu=\nu^{(1)}}^{%
\nu^{(2)}}F_{s}^{(+)}(\beta;\,\nu)\,,\ \ 1\leqslant\nu^{(1)}\leqslant
\nu^{(2)}\leqslant\infty\,.   \label{9}
\end{equation}
It is obvious that the following relations hold true:%
\begin{equation}
\Phi_{s}^{(+)}(\beta)=\widetilde{\Phi}_{s}(\beta;\,\nu^{(1)}=1,\,\nu
^{(2)}=\infty)\,;\ \ \Phi_{s}^{(+)}(\beta)>\widetilde{\Phi}_{s}(\beta
;\,\nu^{(1)}\geqslant1,\,\nu^{(2)}<\infty)\,.   \label{10}
\end{equation}
Let us consider a set of such $\nu^{(1)},\,\,\nu^{(2)}\,(1\leqslant\nu
^{(1)}\leqslant\nu^{(2)}<\infty)$ that satisfy the inequality 
\begin{equation}
\widetilde{\Phi}_{s}(\beta;\,\nu^{(1)},\,\nu^{(2)})\geqslant\frac{1}{2}%
\Phi_{s}^{(+)}(\beta)\,.   \label{11}
\end{equation}
Obviously, such $\nu^{(1)},\,\nu^{(2)}$ do exist for any $\beta$ (for
instance, the case $\nu^{(1)}=1$ provides the existence of such a finite $%
\nu^{(2)}$). It is equally obvious that the condition (\ref{11}) alone is
generally insufficient to determine such a pair of values $%
\nu^{(1)},\,\nu^{(2)}$. Let us choose such $\nu_{s}^{(1)}(\beta),\,\,%
\nu_{s}^{(2)}(\beta)$ that the fulfillment of (\ref{11}) implies that the
difference $\nu_{s}^{(2)}(\beta)-\nu_{s}^{(1)}(\beta)$ is minimal, and the
following non-negative value is also minimal: 
\begin{equation}
\widetilde{\Phi}_{s}(\beta;\,\nu_{s}^{(1)}(\beta),\,\nu_{s}^{(2)}(\beta))-%
\frac{1}{2}\Phi_{s}^{(+)}(\beta)\geqslant0\,.   \label{12}
\end{equation}

A definition equivalent to the above for the effective width of the spectrum
can be given using the concept of the partial contribution $P_{s}(\beta;\nu),
$%
\begin{equation}
P_{s}(\beta;\nu)=\frac{F_{s}^{(+)}(\beta;\,\nu)}{\Phi_{s}^{(+)}(\beta )}\,, 
\label{14}
\end{equation}
of separate harmonics of the spectrum, introduced in \cite{9}. Then (\ref{4}%
) implies the property%
\begin{equation}
\sum_{\nu=1}^{\infty}P_{s}(\beta;\nu)=1\,.   \label{15}
\end{equation}
Let us choose $\nu_{s}^{(1)}(\beta)$ and $\nu_{s}^{(2)}(\beta)$ such that at
the minimum difference $\nu_{s}^{(2)}(\beta)-\nu_{s}^{(1)}(\beta)$ the
following non-negative quantity is minimal:%
\begin{equation}
\sum_{\nu=\nu_{s}^{(1)}(\beta)}^{\nu_{s}^{(2)}(\beta)}P_{s}(\beta;\nu )-%
\frac{1}{2}\geqslant0\,.   \label{16}
\end{equation}
Introducing $\Lambda_{s}(\beta)$ in accordance with (\ref{13}), we arrive at
the following equivalent definition: the effective width of the spectrum is
the minimum spectral range at which the sum of the partial contributions of
separate harmonics is no less than $1/2$.

From a practical point of view, the most interesting case is presented by
the ultrarelativistic limit ($\beta\approx1$, which is equivalent to $%
\gamma\gg1$) of SR. In this case, a big part of the study of the effective
width, as well as the study of other physically interesting quantitative
characteristics of the spectral distribution of SR polarization components,
can be done analytically.

\section{The ultrarelativistic case}

At the higher energies of a charge (which implies the condition $\gamma\gg1$%
), we can use the well-known \cite{1,2,3,4,5} approximations of the Bessel
functions by using the MacDonald functions (Bessel functions of 2nd kind)
and replacing the summation over $\nu$ by integration. As a result, the
expressions (\ref{4}) for $\Phi_{s}^{(+)}(\beta\approx1)=\Phi_{s}^{(+)}$ can
be written in the following form 
\begin{equation}
\Phi_{s}^{(+)}=\int_{0}^{\infty}F_{s}^{(+)}(y)dy\,,\ \ y=\frac{2\nu}{%
3\gamma^{3}}\,,   \label{17}
\end{equation}
where the spectral densities $F_{s}^{(+)}(y)$ (\ref{6}) take the form%
\begin{align}
F_{2}^{(+)}(y) & =\frac{9\sqrt{3}}{32\pi}\,y\left[ 3K_{2/3}(y)-\int
_{y}^{\infty}K_{1/3}(x)dx\right] \,,  \notag \\
F_{3}^{(+)}(y) & =\frac{9\sqrt{3}}{32\pi}\,y\left[ K_{2/3}(y)-\int
_{y}^{\infty}K_{1/3}(x)dx\right] \,,  \notag \\
F_{0}^{(+)}(y) & =F_{2}^{(+)}(y)+F_{3}^{(+)}(y)\,,\ \ F_{\pm1}^{(+)}(y)=%
\frac{1}{2}F_{0}^{(+)}(y)\pm\frac{9}{16\pi^{2}}\,y\,K_{1/3}^{2}\left( \frac{y%
}{2}\right) \,.   \label{18}
\end{align}
Here, $K_{1/3}(x)$ and $K_{2/3}(x)$ are MacDonald functions (Bessel
functions of 2nd kind).

\begin{figure}
\includegraphics[width=\textwidth]{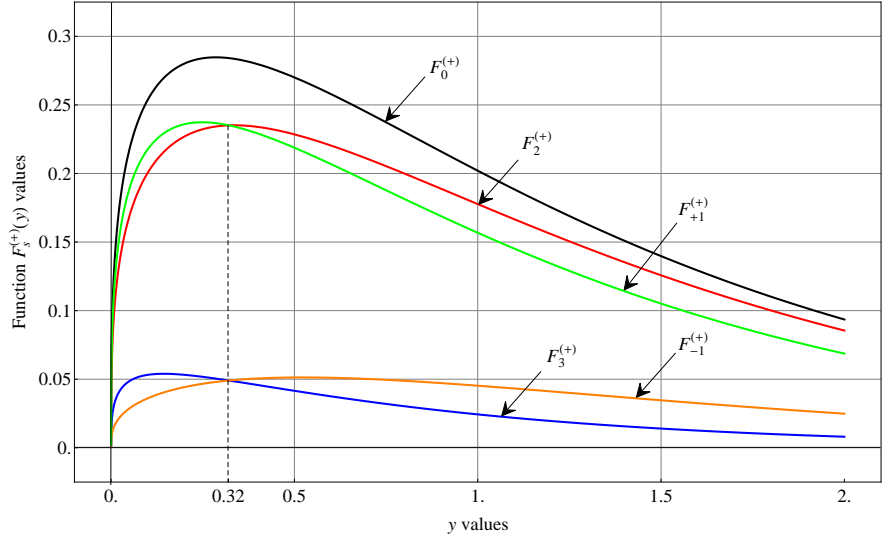} \centering
\caption{Spectral distribution for $F_{s}^{(+)}(y)$.}\label{Fig1}
\end{figure}

The functions $F_{s}^{(+)}(y)$ are plotted in Fig. \ref{Fig1}.

Let us also introduce the function $\Phi_{s}^{(+)}(y)$ 
\begin{equation}
\Phi_{s}^{(+)}(y)=\int_{0}^{y}F_{s}^{(+)}(x)dx\,;\ \ \Phi_{s}^{(+)}=\Phi
_{s}^{(+)}(y=\infty)\,.   \label{19}
\end{equation}
The values of $\Phi_{s}^{(+)}$, according to (\ref{7}) and taking account of
(\ref{8}), are known exactly:%
\begin{align}
\Phi_{2}^{(+)} & =\frac{7}{16}\,,\ \ \Phi_{3}^{(+)}=\frac{1}{16}\,,\ \
\Phi_{0}^{(+)}=\frac{1}{2}\,,  \notag \\
\Phi_{\pm1}^{(+)} & =\frac{1}{4}\left( 1\pm\frac{\sqrt{3}}{\pi}\right)
\approx\frac{1}{2}\left\{ 
\begin{array}{c}
0,7756644 \\ 
0,2243356%
\end{array}
\right. =\left\{ 
\begin{array}{c}
0,3878322 \\ 
0,1121678%
\end{array}
\right. .   \label{20}
\end{align}

In particular, it follows from (\ref{20}), that, in the upper half-space, $%
77.6\%$ of the radiated power is attributed to right circular polarization
and $22.4\%$ is attributed to left circular polarization.

The functions $\Phi_{s}^{(+)}(y)$ can be written down as follows:%
\begin{align}
\Phi_{2}^{(+)}(y) & =\frac{9\sqrt{3}}{32\pi}\,\left[ 3J_{1}(y)+J_{2}(y)-%
\frac{\pi\,y^{2}}{2\sqrt{3}}\right] \,,\ \Phi_{3}^{(+)}(y)=\Phi_{2}^{(+)}(y)+%
\frac{9\sqrt{3}}{16\pi}\,J_{1}(y)\,,  \notag \\
\Phi_{0}^{(+)}(y) & =\Phi_{2}^{(+)}(y)+\Phi_{3}^{(+)}(y)\,,\ \Phi_{\pm
1}^{(+)}(y)=\frac{1}{2}\Phi_{0}^{(+)}(y)\pm\frac{3}{8\pi^{2}}\,J_{3}\left( 
\frac{y}{2}\right) \pm\frac{\sqrt{3}}{4\pi}\,,   \label{21}
\end{align}
where the following notations are used%
\begin{align}
J_{1}(y) & =\frac{2}{3}\int_{0}^{y}K_{1/3}(x)dx-yK_{1/3}(y)\,,  \notag \\
J_{2}(y) & =\left( \frac{y^{2}}{2}-\frac{4}{9}\right)
\int_{0}^{y}K_{1/3}(x)dx+\frac{2}{3}yK_{1/3}(y)+\frac{y^{2}}{2}K_{2/3}(y)\,,
\notag \\
J_{3}(y) & =3y^{2}\left[ K_{1/3}^{2}(y)-K_{2/3}^{2}(y)\right]
-2yK_{1/3}(y)K_{2/3}(y)\,.   \label{22}
\end{align}

\subsection{The effective width of the spectrum in the ultrarelativistic
limit}

According to the given above definition, to determine the effective width of
the spectrum in the ultrarelativistic limit, one has to find such $%
y_{s}^{(1)}$ and $y_{s}^{(2)}>y_{s}^{(1)}$ that 
\begin{equation}
\int_{y_{s}^{(1)}}^{y_{s}^{(2)}}F_{s}^{(+)}(y)dy=\frac{1}{2}%
\Phi_{s}^{(+)}\,,   \label{24}
\end{equation}
and given this, the differences $y_{s}^{(2)}-y_{s}^{(1)}=\Delta_{s}>0$ must
be minimal.

It is convenient to use Eq. (\ref{24}) to determine first $y_{s}^{(2)}$ as a
function of $y_{s}^{(1)}$, or equivalently, to determine $\Delta_{s}$ as a
function $y_{s}^{(1)}$. Let us write%
\begin{equation}
y_{s}^{(2)}\left( y_{s}^{(1)}\right) =y_{s}^{(1)}+\Delta_{s}\left(
y_{s}^{(1)}\right) .  \notag
\end{equation}
This implies the following relation:%
\begin{equation}
y_{s}^{(2)\,^{\prime}}\equiv\frac{dy_{s}^{(2)}\left( y_{s}^{(1)}\right) }{%
dy_{s}^{(1)}}=1+\Delta_{s}^{\prime},\ \ \Delta_{s}^{\prime}\equiv \frac{%
d\Delta_{s}\left( y_{s}^{(1)}\right) }{dy_{s}^{(1)}}\,.   \label{25}
\end{equation}
Taking (\ref{25}) into account, we differentiate (\ref{24}) with respect to $%
y_{s}^{(1)}$ to obtain%
\begin{equation}
\left( 1+\Delta_{s}^{\prime}\right) F_{s}^{(+)}\left( y_{s}^{(2)}\right)
-F_{s}^{(+)}\left( y_{s}^{(1)}\right) =0\,.   \label{26}
\end{equation}
However, we need to find such $y_{s}^{(1)}$ that $\Delta_{s}\left(
y_{s}^{(1)}\right) $ is minimal, which implies that $\Delta_{s}^{\prime}=0$
at the required point. Finally, in accordance with 
\begin{equation}
\int_{y_{s}^{(1)}}^{\,y_{s}^{(2)}}F_{s}^{(+)}(y)dy=\Phi_{s}^{(+)}\left(
y_{s}^{(2)}\right) -\Phi_{s}^{(+)}\left( y_{s}^{(1)}\right) \,,   \label{27}
\end{equation}
we arrive at the following set of equations for the quantities $y_{s}^{(1)}$
and $y_{s}^{(2)}$:%
\begin{equation}
\Phi_{s}^{(+)}\left( y_{s}^{(2)}\right) -\Phi_{s}^{(+)}\left(
y_{s}^{(1)}\right) =\frac{1}{2}\Phi_{s}^{(+)}\,,\ \ F_{s}^{(+)}\left(
y_{s}^{(2)}\right) =F_{s}^{(+)}\left( y_{s}^{(1)}\right) \,,\ \
y_{s}^{(2)}>y_{s}^{(1)}\,,   \label{28}
\end{equation}
which allows one to determine them uniquely. Since, in accordance with (\ref%
{17}), the relation $2\nu=3y\gamma^{3}$ holds true, we obtain in the
ultrarelativistic limit 
\begin{align}
& \nu_{s}^{(1)}(\beta)=a_{s}^{(1)}\gamma^{3},\ \ \nu_{s}^{(2)}(\beta
)=a_{s}^{(2)}\gamma^{3},  \notag \\
& \Lambda_{s}(\beta)=b_{s}\gamma^{3},\ \ b_{s}=a_{s}^{(2)}-a_{s}^{(1)},\ \
a_{s}^{(k)}=\frac{3}{2}y_{s}^{(k)}.   \label{29}
\end{align}

The distribution of the harmonics $\nu_{s}^{(1)}(\beta)$ and $%
\nu_{s}^{(2)}(\beta)$ and the quantities $\Lambda_{s}(\beta)$ follows
exactly the same order. The effective width of the spectrum $%
\Lambda_{s}(\beta)$ is different for different polarizations, for instance,
the ratio of the largest quantity $\Lambda_{-1}(\beta)$ to the smallest
quantity $\Lambda_{3}(\beta)$ is $\sim$ $1,76$.

For each type of polarization (each $s$) the frequencies $\nu<\nu_{s}^{(max)}
$ will be referred to as low, and the frequencies $\nu>\nu_{s}^{(max)}$ will
be referred to as high. It is obvious that the portion $r_{s}^{(1)}$ of the
effective width of the spectrum which corresponds to the low frequencies is
given by 
\begin{equation}
r_{s}^{(1)}=\frac{\nu_{s}^{(max)}(\beta)-\nu_{s}^{(1)}(\beta)}{%
\nu_{s}^{(2)}(\beta)-\nu_{s}^{(1)}(\beta)}=\frac{a_{s}^{(max)}-a_{s}^{(1)}}{%
b_{s}}\,.   \label{32}
\end{equation}

\subsection{Another characteristics of spectral distributions in the
ultrarelativistic limit}

The expressions (\ref{18}) and (\ref{21}) make it possible to obtain exact
quantitative data for interesting physical characteristics of spectral
distributions. These quantitative data are presented in the following table %
\ref{Table}.

Calculating the location of the maximum points $y_{s}^{(max)}$ of spectral
distributions, one determines the frequencies that correspond to the maxima: 
\begin{equation}
\nu_{s}^{(max)}(\beta)=a_{s}^{(max)}\gamma^{3}\,,\ \ a_{s}^{(max)}=\frac{3}{2%
}y_{s}^{(max)}.   \label{30}
\end{equation}
The numerical values $a_{s}^{(max)}$ for $s=0,2,3$ are already known,
whereas those for $s=\pm1$ are presented here for the first time. The
frequencies corresponding to the maxima are distributed in the following
order for all values of $\beta$:%
\begin{equation}
\nu_{3}^{(max)}(\beta)<\nu_{1}^{(max)}(\beta)<\nu_{0}^{(max)}(\beta)<\nu
_{2}^{(max)}(\beta)<\nu_{-1}^{(max)}(\beta)\,.   \label{31}
\end{equation}

\begin{table}
\centering
\resizebox{\columnwidth}{!}{\begin{tabular}{|c|c|c|c|c|c|}
\hline
s & $0$ & $2$ & $3$ & $-1$ & $+1$\tabularnewline
\hline
\hline
$y_{s}^{(max)}$ & $2.85812e-01$ & $3.35524e-01$ & $1.43921e-01$ & $5.22405e-01$ & $2.48583e-01$\tabularnewline
\hline
$F_{s}^{(+)}\left(y_{s}^{(max)}\right)$ & $2.84696e-01$ & $2.35158e-01$ & $5.39423e-02$ & $5.12872e-02$ & $2.37335e-01$\tabularnewline
\hline
$\Phi_{s}^{(+)}\left(y_{s}^{(max)}\right)$ & $7.17052e-02$ & $6.90125e-02$ & $6.90380e-03$ & $2.26361e-02$ & $5.21376e-02$\tabularnewline
\hline
$\eta_{s}^{(max)}$ & $1.43410e-01$ & $1.57743e-01$ & $1.10461e-01$ & $2.01806e-01$ & $1.34433e-01$\tabularnewline
\hline
$y_{s}^{(1)}$ & $3.49398e-02$ & $4.87043e-02$ & $1.08505e-02$ & $1.22065e-01$ & $2.71081e-02$\tabularnewline
\hline
$F_{s}^{(+)}\left(y_{s}^{(1)}\right)$ & $1.98326e-01$ & $1.67772e-01$ & $3.39459e-02$ & $3.84141e-02$ & $1.61750e-01$\tabularnewline
\hline
$\Phi_{s}^{(+)}\left(y_{s}^{(1)}\right)$ & $5.36798e-03$ & $6.31493e-03$ & $2.84108e-04$ & $3.52777e-03$ & $3.39459e-03$\tabularnewline
\hline
$\eta_{s}^{(1)}$ & $1.0736e-02$ & $1.44341e-02$ & $4.54573e-03$ & $3.14508e-02$ & $8.75273e-03$\tabularnewline
\hline
$y_{s}^{(2)}$ & $1.02680e+00$ & $1.08939e+00$ & $6.94023e-01$ & $1.32370e+00$ & $9.58312e-01$\tabularnewline
\hline
$\Phi_{s}^{(+)}\left(y_{s}^{(2)}\right)$ & $2.55368e-01$ & $2.25065e-01$ & $3.15341e-02$ & $5.96117e-02$ & $1.97311e-01$\tabularnewline
\hline
$\eta_{s}^{(2)}$ & $5.10736e-01$ & $5.14434e-01$ & $5.04546e-01$ & $5.31451e-01$ & $5.08753e-01$\tabularnewline
\hline
$y_{s}^{(3)}$ & $1.10709e-02$ & $1.44604e-02$ & $4.90942e-03$ & $3.59457e-02$ & $9.26077e-03$\tabularnewline
\hline
$\Phi_{s}^{(+)}\left(y_{s}^{(3)}\right)$ & $1.19916e-03$ & $1.29074e-03$ & $1.00954e-04$ & $6.86800e-04$ & $8.36754e-04$\tabularnewline
\hline
$\eta_{s}^{(3)}$ & $2.39832e-03$ & $2.95025e-03$ & $1.61526e-03$ & $6.12297e-03$ & $2.15752e-03$\tabularnewline
\hline
$y_{s}^{(4)}$ & $1.47628e+00$ & $1.59002e+00$ & $9.06361e-01$ & $1.95582e+00$ & $1.35291e+00$\tabularnewline
\hline
$\Phi_{s}^{(+)}\left(y_{s}^{(4)}\right)$ & $3.31467e-01$ & $2.96035e-01$ & $3.79763e-02$ & $7.97257e-02$ & $2.52321e-01$\tabularnewline
\hline
$\eta_{s}^{(4)}$ & $6.62933e-01$ & $6.76652e-01$ & $6.07621e-01$ & $7.10772e-01$ & $6.50593e-01$\tabularnewline
\hline
$a_{s}^{(max)}$ & $4.28718e-01$ & $5.03287e-01$ & $2.15881e-01$ & $7.83608e-01$ & $3.72875e-01$\tabularnewline
\hline
$a_{s}^{(1)}$ & $5.24096e-02$ & $7.30564e-02$ & $1.62757e-02$ & $1.83097e-01$ & $4.06621e-02$\tabularnewline
\hline
$a_{s}^{(2)}$ & $1.54021e+00$ & $1.63408e+00$ & $1.04103e+00$ & $1.98555e+00$ & $1.43747e+00$\tabularnewline
\hline
$a_{s}^{(3)}$ & $1.66063e-02$ & $2.16906e-02$ & $7.36413e-03$ & $5.39186e-02$ & $1.38912e-02$\tabularnewline
\hline
$a_{s}^{(4)}$ & $2.21442e+00$ & $2.38502e+00$ & $1.35954e+00$ & $2.93372e+00$ & $2.02936e+00$\tabularnewline
\hline
$b_{s}$ & $1.4878e+00$ & $1.56103e+00$ & $1.02476e+00$ & $1.80245e+00$ & $1.39681e+00$\tabularnewline
\hline
$d_{s}$ & $2.19781e+00$ & $2.36333e+00$ & $1.35218e+00$ & $2.87981e+00$ & $2.01547e+00$\tabularnewline
\hline
$r_{s}^{(1)}$ & $2.52929e-01$ & $2.75607e-01$ & $1.94782e-01$ & $3.33163e-01$ & $2.37837e-01$\tabularnewline
\hline
$r_{s}^{(2)}$ & $1.32674e-01$ & $1.43309e-01$ & $1.05915e-01$ & $1.70355e-01$ & $1.25680e-01$\tabularnewline
\hline
$r_{s}^{(3)}$ & $6.60535e-01$ & $6.73701e-01$ & $6.06006e-01$ & $7.04649e-01$ & $6.48435e-01$\tabularnewline
\hline
\end{tabular}
}\caption{Spectral emission characteristics in the ultra relativistic limit.}%
\label{Table}
\end{table}

The quantity $100\,r_{s}^{(1)}$ determines the percentage of the effective
width of the spectrum that is contributed by the low frequencies for each
polarization. As we can see, the quantities $r_{s}^{(1)}$ are distributed in
the order (\ref{31}). It is quite interesting that the low frequencies
contribute to a significantly smaller part of the effective width of the
spectrum (from the maximum part $\sim$ $33,3\%$ for $s=-1$ to the minimum
part $\sim$ $19,5\%$ for $s=3$), and therefore, most of the effective width
of the spectrum is contributed by the high frequencies.

Let us introduce the quantities 
\begin{equation}
\eta_{s}^{(k)}=\frac{\Phi_{s}^{(+)}(y_{s}^{(k)})}{\Phi_{s}^{(+)}}\,, 
\label{33}
\end{equation}
that determine, for each $s$-component of SR polarization, the portion of
power radiated by the interval of frequencies $0<\nu<\nu_{s}^{(k)}$ from the
total power emitted at this polarization. It is obvious that the quantity $%
100\,\eta_{s}^{(k)}$ accordingly determines the percentage of power radiated
at the spectral region $0<\nu<\nu_{s}^{(k)}$. The order of distribution of $%
\eta_{s}^{(k)}$ is identical with (\ref{31}).

The numerical values $\eta_{s}^{(max)}$ provide convincing evidence of the
fact that a substantially larger part of the radiated power at each
component of polarization is contributed by the region of high frequencies
(the maximum percentage of radiated power at low frequencies is $\sim$ $%
20,2\%$ for $s=-1$ and the minimum is $\sim$ $11,0\%$ for $s=3$, and
therefore from $\sim$ $80\%$ to $\sim$ $90\%$ of the radiated power is
contributed by the high frequencies).

The numerical values $\eta_{s}^{(1)}$ represent the portion of radiated
power in the frequency range up to the beginning of the effective width of
the spectrum. This portion is quite insignificant (the maximum percentage
being $\sim$ $3,15\%$ for $s=-1$ and the minimum being $\sim$ $0,45\%$ for $%
s=3$).

The values $r_{s}^{(2)}=\eta_{s}^{(max)}-\eta_{s}^{(1)}$ determine the
portion of radiated power at low frequencies that corresponds to the
effective width of the spectrum (here, the maximum percentage of power
radiated at low frequencies is $\sim$ $17,0\%$ for $s=-1$ and the minimum
percentage is $\sim$ $10,6\%$ for $s=3$; there is an approximation $%
r_{s}^{(1)}\sim2r_{s}^{(2)}$, which is physically quite justified, since the
radiated powers are approximately proportional to the frequency ranges).

In optics one characterizes the profiles of spectral distributions by
introducing the concept of the half-width of the spectrum. Such a quantity
can also be introduced in the case under consideration.

Let us find some pairs of points $y_{s}^{(3)}\,,y_{s}^{(4)}$ related by%
\begin{equation}
F_{s}^{(+)}(y_{s}^{(3)})=F_{s}^{(+)}(y_{s}^{(4)})=\frac{1}{2}%
F_{s\,max}^{(+)}\,,\ \ y_{s}^{(3)}<y_{s}^{(4)}\,,   \label{34}
\end{equation}
and calculate for these quantities characteristics similar to those
considered above. In particular, the half-width of the spectrum $%
\Lambda_{s}^{1/2}(\beta)$ is determined by%
\begin{equation}
\Lambda_{s}^{1/2}(\beta)=d_{s}\gamma^{3}\,,\ \
d_{s}=a_{s}^{(4)}-a_{s}^{(3)}\,,\ \ a_{s}^{(k)}=\frac{3}{2}y_{s}^{(k)}, 
\label{35}
\end{equation}
and the portion of radiated power that is attributed to the half-width is
determined by the parameter $r_{s}^{(3)}=\eta_{s}^{(4)}-\eta_{s}^{(3)}$.

As we can see from the table, the half-width is attributed to the range from 
$\sim$ $60\%$ to $\sim$ $70\%$ of radiated power, and the remaining part is
radiated in the region of high frequencies (up to the beginning of the
half-width, the radiated power is merely in the range of $\sim$ $0,16\%$ to $%
\sim$ $0,61\%$). Consequently, for spectral distributions of SR the concept
of the half-width of the spectrum is less informative than the concept of
the effective width of the spectrum that we have proposed in this article.

\section{Brief summary}

We give a new definition of the effective width for the SR spectrum and
calculate the effective width for all polarization components of the SR in
the ultra-relativistic limit. For the first time, the spectral distribution
for the circular polarization component of the SR for the upper half-plane
is obtained within classical theory. In addition, the relative radiation
power emitted in some physically interesting spectral ranges is found. 

\section{Acknowledgments}
Bagrov thanks FAPESP for its support and IF USP for its hospitality. The
reported study of Gitman was partially supported by RFBR, research project
No. 15- 02-00293a. Gitman thanks CNPq and FAPESP for their permanent
support. The work of Loginov is partially supported by the Ministry of
Science of the Russian Federation (grant N\textsuperscript{\underline{o}}
2014/223), Code project 1766.

\appendix

\section{Relevant equations of the synchrotron radiation theory}

We present here some well-known in classical theory equations that describe
the physical characteristics of synchrotron radiation (SR), see for example,
the books \cite{1,2,3,4,5}.

The spectral-angular distribution of the radiation power of SR polarization
components can be presented in the form%
\begin{equation}
W_{s}=W\sum_{\nu=1}^{\infty}\int_{0}^{\pi}f_{s}(\beta;\,\nu,\,\theta
)\sin\theta d\theta.   \label{1}
\end{equation}
Here, $\theta$ is the angle between the direction of the magnetic field and
the momentum of the field of radiation; $\nu$ is the number of emitted
harmonics; the velocity of the charge in orbit is $v=c\beta$ , where $c$ is
the speed of light; $W$ is the total radiated power of unpolarized
radiation, which can be written down in the form%
\begin{equation}
W=\frac{2}{3}\frac{ce^{2}}{R^{2}}(\gamma^{2}-1)^{2}=\frac{2}{3}\frac {%
e^{4}H^{2}(\gamma^{2}-1)}{m_{0}^{2}c^{3}},\ \ \gamma=\frac{1}{\sqrt {%
1-\beta^{2}}}\,,   \label{2}
\end{equation}
where $e$ is the charge of the particle; $R$ is the radius of the orbit; $H$
is the strength of the magnetic field; $m_{0}$ is the rest mass of the
charged particle; $\gamma$ is the relativistic factor. The index $s$ serves
to number the polarization components: $s=2$ corresponds to the $\sigma$%
-component of linear polarization; $s=3$ corresponds to the $\pi$-component
of linear polarization; $s=1$ stands for the right circular polarization; $%
s=-1$ stands for the left circular polarization; $s=0$ corresponds to the
power of unpolarized radiation. The functions $f_{s}(\beta;\,\nu,\,\theta)$
have the form 
\begin{align}
f_{2}(\beta;\,\nu,\,\theta) & =\frac{3\nu^{2}}{2\gamma^{4}}J_{\nu}^{\prime
}\,^{2}(x);\ \ f_{3}(\beta;\,\nu,\,\theta)=\frac{3\nu^{2}}{2\gamma^{4}}\frac{%
\cos^{2}\theta}{\beta^{2}\sin^{2}\theta}J_{\nu}^{2}(x);  \notag \\
f_{\pm1}(\beta;\,\nu,\,\theta) & =\frac{3\nu^{2}}{4\gamma^{4}}\left[ J_{\nu
}^{\prime}(x)\pm\varepsilon\frac{\cos\theta}{\beta\sin\theta}J_{\nu }(x)%
\right] ^{2};\ \ x=\nu\beta\sin\theta;\ \ \varepsilon=-\frac{e}{|e|};  \notag
\\
f_{0}(\beta;\,\nu,\,\theta) & =f_{2}(\beta;\,\nu,\,\theta)+f_{3}(\beta
;\,\nu,\,\theta)=f_{1}(\beta;\,\nu,\,\theta)+f_{-1}(\beta;\,\nu,\,\theta
)\,,   \label{3}
\end{align}
where, $J_{\nu}(x)$ are the Bessel functions. The case of an electron
corresponds to $\varepsilon=1$.

It is well known that in the region $0\leqslant\theta<\pi/2$ (this region
will be called the upper half-space) the right circular polarization is
dominant, whereas in the region $\pi/2<\theta\leqslant\pi$ (this region will
be called the lower half-space) it is the left circular polarization that is
dominant (exact quantitative characteristics of this property of SR were
first obtained in \cite{6,7}). However, if in expression (\ref{1}) one
integrates over $\theta\ \ (0\leqslant\theta\leqslant\pi)$ then the
differences in the spectral distribution of the right and left circular
polarizations disappear. To identify these differences, we present the
expressions (\ref{1}) in the form 
\begin{align}
& W_{s}=W\left[ \Phi_{s}^{(+)}(\beta)+\Phi_{s}^{(-)}(\beta)\right] \,;\ \
\Phi_{s}^{(+)}(\beta)=\sum_{\nu=1}^{\infty}F_{s}^{(+)}(\beta ;\,\nu)\,,\ \
\Phi_{s}^{(-)}(\beta)=\sum_{\nu=1}^{\infty}F_{s}^{(-)}(\beta;\,\nu)\,; 
\notag \\
& F_{s}^{(+)}(\beta;\,\nu)=\int_{0}^{\pi/2}f_{s}(\beta;\,\nu,\,\theta
)\sin\theta d\theta\,,\ \
F_{s}^{(-)}(\beta;\,\nu)=\int_{\pi/2}^{\pi}f_{s}(\beta;\,\nu,\,\theta)\sin%
\theta d\theta\,,   \label{4}
\end{align}
and, besides, it is sufficient to study the properties of the functions $%
F_{s}^{(+)}(\beta;\,\nu)$ (respectively, the properties of the functions $%
\Phi_{s}^{(+)}(\beta)$), since there exist the obvious relations 
\begin{align}
F_{s}^{(-)}(\beta;\,\nu) & =F_{s}^{(+)}(\beta;\,\nu)\,,\ \
\Phi_{s}^{(-)}(\beta)=\Phi_{s}^{(+)}(\beta)\,,\ \ s=0,\,2,\,3;  \notag \\
F_{1}^{(-)}(\beta;\,\nu) & =F_{-1}^{(+)}(\beta;\,\nu)\,,\ \
\Phi_{1}^{(-)}(\beta)=\Phi_{-1}^{(+)}(\beta)\,,  \notag \\
F_{-1}^{(-)}(\beta;\,\nu) & =F_{1}^{(+)}(\beta;\,\nu)\,,\ \
\Phi_{-1}^{(-)}(\beta)=\Phi_{1}^{(+)}(\beta)\,.   \label{5}
\end{align}
Integration over $\theta$ in the upper half-space $0\leqslant\theta
\leqslant\pi/2$ in (\ref{4}) can be carried out exactly, which leads to the
following expressions:%
\begin{align}
F_{2}^{(+)}(\beta;\,\nu) & =\frac{3\nu}{4\,\gamma\,^{4}\beta^{3}}\left[
2\beta^{2}J_{2\nu}^{\prime}(2\nu\beta)+\beta^{2}\int_{0}^{2\nu\beta}J_{2\nu
}(x)dx-2\nu\beta\int_{0}^{2\nu\beta}\frac{J_{2\nu}(x)}{x}dx\right] \,, 
\notag \\
F_{3}^{(+)}(\beta;\,\nu) & =\frac{3\nu}{4\,\gamma\,^{4}\beta^{3}}\left[
2\nu\beta\int_{0}^{2\nu\beta}\frac{J_{2\nu}(x)}{x}dx-\int_{0}^{2\nu\beta
}J_{2\nu}(x)dx\right] \,,  \notag \\
F_{0}^{(+)}(\beta;\,\nu) & =F_{2}^{(+)}(\beta;\,\nu)+F_{3}^{(+)}(\beta
;\,\nu)=\frac{3\nu}{4\,\gamma\,^{4}\beta^{3}}\left[ 2\beta^{2}J_{2\nu
}^{\prime}(2\nu\beta)-(1-\beta^{2})\int_{0}^{2\nu\beta}J_{2\nu}(x)dx\right]
\,,  \notag \\
F_{\pm1}^{(+)}(\beta;\,\nu) & =\frac{1}{2}F_{0}^{(+)}(\beta;\,\nu)\pm \frac{%
3\nu J\,_{\nu}^{2}(\nu\beta)}{4\,\gamma\,^{4}\beta^{2}}\,.   \label{6}
\end{align}
The sums over the harmonics $\nu$ in (\ref{4}) can also be calculated
exactly, which leads to the expressions%
\begin{equation}
\Phi_{2}^{(+)}(\beta)=\frac{6+\beta^{2}}{16}\,,\ \ \Phi_{3}^{(+)}(\beta )=%
\frac{2-\beta^{2}}{16}\,,\ \ \Phi_{0}^{(+)}(\beta)=\frac{1}{2}\,,\ \
\Phi_{\pm1}^{(+)}(\beta)=\frac{1}{4}\left[ 1\pm\frac{3}{4}\chi _{1}(\beta)%
\right] \,.   \label{7}
\end{equation}
The function $\chi_{1}(\beta)$ introduced here has been defined and studied
in \cite{7}. In particular, the study of \cite{7} demonstrated that in the
segment $0\leqslant\beta\leqslant1$ the function $\chi_{1}(\beta)$ is finite
and decreases monotonously; at the ends of this segment, it takes the
following values:%
\begin{equation}
\chi_{1}(0)=1\,,\ \ \chi_{1}(1)=\frac{4}{\pi\sqrt{3}}\,.   \label{8}
\end{equation}

\end{document}